\documentclass{PoS}

\usepackage{cite}
\usepackage{amsmath, amsthm, amssymb}
\usepackage{url}
\usepackage{graphicx}

\newcommand{\MS}{\overline{\text{MS}}}
\newcommand{\ep}{\varepsilon}
\allowdisplaybreaks[4]

\title{{\footnotesize 
       DESY 19--187, DO-TH 19/26, SAGEX-19-27, PoS(RADCOR19)047}\\
The Polarized Three-Loop Anomalous Dimensions from a Massive Calculation}

\ShortTitle{Polarized anomalous dimensions}

\author{Arnd Behring \\
      Institut f\"ur Theoretische Teilchenphysik, Campus S\"ud, 
      Karlsruher Institut f\"ur Technologie (KIT), D-76128, Germany. \\
      E-mail: \email{arnd.behring@kit.edu}
}

\author{\speaker{Johannes Bl{\"u}mlein} \\
      Deutsches Elektronen--Synchrotron, DESY,
      Platanenallee 6, D-15738 Zeuthen, Germany.\\
      E-mail: \email{johannes.bluemlein@desy.de}
}

\author{Abilio De Freitas \\
      Deutsches Elektronen--Synchrotron, DESY,
      Platanenallee 6, D-15738 Zeuthen, Germany.\\
      E-mail: \email{abilio.de.freitas@desy.de}
}

\author{Alexander Goedicke \\
      Institut f\"ur Theoretische Teilchenphysik, Campus S\"ud, 
      Karlsruher Institut f\"ur Technologie (KIT), D-76128, Germany. \\
      E-mail: \email{alexander.goedicke@posteo.de}
}

\author{Sebastian Klein \\
Institut f\"ur Theoretische Teilchenphysik und Kosmologie, RWTH Aachen,      
Sommerfeldstr. 16, D-52074 Aachen, Germany.\\
E-mail: \email{sebastian.klein@mail.de}
}

\author{Andreas van Manteuffel \\
      Department of Physics and Astronomy ,
      Michigan State University, East Lansing, MI 48824, USA. \\
      E-mail: \email{manteuffel@pa.msu.edu}
}

\author{Carsten Schneider \\
      Johannes Kepler Universit{\"a}t Linz,
      Altenbergerstraße 69, A-4040 Linz, Austria.\\
      E-mail: \email{clemens.raab@jku.at}
}

\author{Kay Sch{\"o}nwald \\
      Deutsches Elektronen--Synchrotron, DESY,
      Platanenallee 6, D-15738 Zeuthen, Germany.\\
      E-mail: \email{kay.schoenwald@desy.de}
}

\abstract{We present results on the calculation of the polarized 2- and 3-loop anomalous dimensions in a massive 
computation of the associated operator matrix element. We also discuss the treatment of $\gamma_5$ and derive
results in the M-scheme.}

\FullConference{14th International Symposium on Radiative Corrections (RADCOR2019)\\ 
9-13 September 2019\\
		Palais des Papes, Avignon, France}

\begin{document}
%--------------------------------------------------------------------------------------------------------
\section{Introduction} 
\label{sec:1}
%--------------------------------------------------------------------------------------------------------

\vspace*{1mm}
\noindent
The next-to-next-to leading order anomalous dimensions $\gamma_{ij}^{(2)}$ and
splitting functions $P_{ij}^{(2)}$ govern the scale evolution of the polarized
parton distributions in Quantum Chromodynamics to order $\alpha_s^3$.
They are important ingredients for precision measurements at $e \, p$ and 
hadron colliders, different fixed-target experiments and future experiments
like the EIC \cite{FUTURE} and at RHIC.
Furthermore they are instrumental for the precise measurement of the strong
coupling constant $\alpha_s(M_Z)$ \cite{alphas} at these facilities and key processes
like the polarized Drell-Yan cross section, jet production cross sections
and the analysis of deep-inelastic scattering data \cite{PDF}.
Precision analyses of polarized processes are also important to resolve the 
spin-composition of polarized nucleons \cite{Lampe:1998eu,Deur:2018roz}.

The calculation of the 2-loop polarized splitting functions has been performed
already in 1995 \cite{Mertig:1995ny,SP_PS1}. 
The complete 3-loop results have been obtained in \cite{Moch:2014sna} within the so-called M-scheme.
In the flavor non-singlet case the 3-loop splitting function
$P_{qq}^{(2),\text{NS}}$ are the same as in the unpolarized case \cite{Moch:2004pa} and 
have also been obtained in \cite{Ablinger:2014vwa}.
The unpolarized splitting functions have been obtained in \cite{Moch:2004pa,Vogt:2004mw} and
the terms $\propto T_F$ have been confirmed in independent massive
calculations in \cite{Ablinger:2014vwa,Ablinger:2014lka,Ablinger:2014nga,Ablinger:2010ty,Ablinger:2017tan}.
In the present paper we review the corresponding calculation in the 
polarized case. For the complete results see Ref.~ \cite{Behring:2019tus}.
Already in 2010 the odd moments $N=1$--$7$ of the polarized OMEs $A_{Qg}^{(3)}$
and $A_{qg,Q}^{(3)}$ and in 2013 also for $A_{gg,Q}^{(3)}$ have been calculated.
Most recently we also added the moment $N=9$.
The set of moments remained unpublished since an important detail in the
definition of the Larin scheme had to be understood first.

The paper is organized as follows.
In Section~\ref{sec:2}~we review the general structure of the massive OMEs 
and their evaluation using the Larin scheme.
The methods used for the calculation of the massive OMEs are described in Section~\ref{sec:3}.
Section~\ref{sec:4}~contains the conclusions. 
%--------------------------------------------------------------------------------------------------------
\section{The polarized massive Operator Matrix Elements} 
\label{sec:2}
%--------------------------------------------------------------------------------------------------------

\vspace*{1mm}
\noindent
There are seven massive operator matrix elements $A_{qq,Q}^{(3),\text{NS}}$, $A_{qq,Q}^{(3),\text{PS}}$, $A_{Qq}^{(3),\text{PS}}$, $A_{qg,Q}^{(3)}$, $A_{Qg}^{(3)}$, $A_{gq,Q}^{(3)}$ and $A_{gg,Q}^{(3)}$.
The non-singlet OME has already been calculated in \cite{Ablinger:2014vwa}. Due to a 
Ward identity it has to agree in the polarized and unpolarized case and
can be given in the $\MS$ scheme.
The principle structure of the OMEs in Mellin-$N$ space has been given in
Ref.~\cite{Bierenbaum:2009mv}, i.e. one obtains
%---------------------------------------------------------------------------------
\begin{eqnarray}
   \hat{\hat{A}}_{Qg}^{(3)}&=&
                  \Bigl(\frac{\hat{m}^2}{\mu^2}\Bigr)^{3\ep/2}
                     \Biggl[
           \frac{\hat{\gamma}_{qg}^{(0)}}{6\ep^3}
             \Biggl(
                   (N_F+1)\gamma_{gq}^{(0)}\hat{\gamma}_{qg}^{(0)}
                 +\gamma_{qq}^{(0)}
                                \Bigl[
                                        \gamma_{qq}^{(0)}
                                      -2\gamma_{gg}^{(0)}
                                      -6\beta_0
                                      -8\beta_{0,Q}
                                \Bigr]
                 +8\beta_0^2
\nonumber\\ &&
                 +28\beta_{0,Q}\beta_0
                 +24\beta_{0,Q}^2
                  +\gamma_{gg}^{(0)}
                                \Bigl[
                                        \gamma_{gg}^{(0)}
                                       +6\beta_0
                                       +14\beta_{0,Q}
                                \Bigr]
             \Biggr)
          +\frac{1}{6\ep^2}
             \Biggl(
                   \hat{\gamma}_{qg}^{(1)}
                      \Bigl[
                              2\gamma_{qq}^{(0)}
                             -2\gamma_{gg}^{(0)}
                             -8\beta_0
\nonumber\\ &&
                             -10\beta_{0,Q}
                      \Bigr]
                  +\hat{\gamma}_{qg}^{(0)}
                      \Bigl[
                              \hat{\gamma}_{qq}^{(1), {\sf PS}}\{1-2N_F\}
                             +\gamma_{qq}^{(1), {\sf NS}}
                             +\hat{\gamma}_{qq}^{(1), {\sf NS}}
                             +2\hat{\gamma}_{gg}^{(1)}
                             -\gamma_{gg}^{(1)}
                             -2\beta_1
                             -2\beta_{1,Q}
                      \Bigr]
\nonumber\\
&&
                  + 6 \delta m_1^{(-1)} \hat{\gamma}_{qg}^{(0)}
                      \Bigl[
                              \gamma_{gg}^{(0)}
                             -\gamma_{qq}^{(0)}
                             +3\beta_0
                             +5\beta_{0,Q}
                      \Bigr]
             \Biggr)
          +\frac{1}{\ep}
             \Biggl(
                   \frac{\hat{\gamma}_{qg}^{(2)}}{3}
                  -N_F \frac{\hat{\tilde{\gamma}}_{qg}^{(2)}}{3}
                  +\hat{\gamma}_{qg}^{(0)}\Bigl[
                                    a_{gg,Q}^{(2)}
\nonumber\\
&&                                    -N_Fa_{Qq}^{(2),{\sf PS}}
                                          \Bigr]
                  +a_{Qg}^{(2)}
                      \Bigl[
                              \gamma_{qq}^{(0)}
                             -\gamma_{gg}^{(0)}
                             -4\beta_0
                             -4\beta_{0,Q}
                      \Bigr]
                  +\frac{\hat{\gamma}_{qg}^{(0)}\zeta_2}{16}
                      \Bigl[
                              \gamma_{gg}^{(0)} \Bigl\{
                                                        2\gamma_{qq}^{(0)}
                                                       -\gamma_{gg}^{(0)}
                                                       -6\beta_0
\nonumber\\ &&
                                                       +2\beta_{0,Q}
                                                \Bigr\}  
                             -(N_F+1)\gamma_{gq}^{(0)}\hat{\gamma}_{qg}^{(0)}
                             +\gamma_{qq}^{(0)} \Bigl\{
                                                       -\gamma_{qq}^{(0)}
                                                       +6\beta_0
                                                \Bigr\}
                             -8\beta_0^2
                             +4\beta_{0,Q}\beta_0
                             +24\beta_{0,Q}^2
                      \Bigr]
\nonumber\\ &&
                  + \frac{\delta m_1^{(-1)}}{2}
                      \Bigl[
                              -2\hat{\gamma}_{qg}^{(1)}
                              +3\delta m_1^{(-1)}\hat{\gamma}_{qg}^{(0)}
                              +2\delta m_1^{(0)}\hat{\gamma}_{qg}^{(0)}
                      \Bigr]
                  + \delta m_1^{(0)}\hat{\gamma}_{qg}^{(0)}
                       \Bigl[
                               \gamma_{gg}^{(0)}
                              -\gamma_{qq}^{(0)}
                              +2\beta_0
                              +4\beta_{0,Q}
                      \Bigr]
\nonumber\\ &&
                  -\delta m_2^{(-1)}\hat{\gamma}_{qg}^{(0)}
             \Biggr)
                 +a_{Qg}^{(3)}
                  \Biggr]. 
\label{AhhhQg3} 
\end{eqnarray}
%------------------------------------------------------------------------------------------------------------------
In this expression all dependencies on $N$ have been dropped for brevity.
$\mu$ denotes the factorization and renormalization scale, 
$\bar{m}$ the bare heavy quark mass, 
$\ep=D-4$ the dimensional regulator,
$\zeta_l, 
l \in \mathbb{N}, l \geq 2$ the values of the Riemann $\zeta$ function at integer argument. 
$\beta_i$ are the expansion coefficients of the QCD $\beta$-function, 
$\beta_{i,Q}$ are related expansion 
coefficients associated to heavy quark effects, 
$\gamma_{ij}^{(k)}$ the expansion coefficients of the anomalous 
dimensions, and $\delta m_k^{(l)}$ the expansion coefficients of the unrenormalized quark mass. 
The above quantities depend on the color factors $C_A = N_C, C_F = (N_C^2-1)/(2 N_C), T_F = 1/2$ for $SU(N_C)$ and 
$N_C= 3$ for QCD, cf. e.g. Ref.~\cite{Bierenbaum:2009mv} and the number of 
massless quark flavors $N_F$. The coefficients $a_{ij}^{(k)}$ 
denote the constant terms of the OMEs at $k$--loop order and $\bar{a}_{ij}^{(k)}$ the corresponding terms 
at $O(\varepsilon)$, 
cf.~\cite{Buza:1996xr,Klein:2009ig,Blumlein:2019zux,Hasselhuhn:2013swa,POL19,PVFNS}.
Furthermore, we use 
%---------------------------------------------------------------------------------
\begin{eqnarray}\label{eq:r1}
\hat{f}(N_F)   &=&  f(N_F+1) - f(N_F)   \\
\label{eq:r2}
\tilde{f}(N_F) &=&  \frac{\displaystyle f(N_F)}{\displaystyle N_F}~.
\end{eqnarray}
%---------------------------------------------------------------------------------
From the poles $O(1/\ep^3)$ one can obtain the one-loop anomalous dimensions,
from the poles $O(1/\ep^2)$ the full two-loop anomalous dimensions while
from the poles $O(1/\ep)$ the contributions $\propto T_F$ of the three-loop
anomalous dimensions can be extracted.

We work in the Larin scheme to describe $\gamma_5$ in $D$ dimensions \cite{Larin:1993tq}. 
In this scheme $\gamma^5$ is described by
%---------------------------------------------------------------------------------
\begin{eqnarray}
\gamma^5 &=& \frac{i}{24} \ep_{\mu \nu \rho \sigma} \gamma^\mu \gamma^\nu \gamma^\rho 
\gamma^\sigma,
\\
\Delta \hspace*{-2.5mm} \slash ~\gamma^5 &=& \frac{i}{6} \ep_{\mu \nu \rho \sigma} \Delta^\mu \gamma^\nu \gamma^\rho 
\gamma^\sigma.
\end{eqnarray}
%---------------------------------------------------------------------------------
and two Levi-Civita symbols are contracted in $D$ dimensions using
%---------------------------------------------------------------------------------
\begin{eqnarray}
\label{eq:epcontr}
\ep_{\mu \nu \rho \sigma}  \ep^{\alpha \lambda  \tau  \gamma} = - {\rm Det}[g_\omega^\beta], 
~~~~\beta = \alpha, \lambda,  \tau,  \gamma;~~\omega = \mu, \nu, \rho, \sigma.
\end{eqnarray}
%---------------------------------------------------------------------------------
We use the projectors $P_g$ and $P_q$ for the amplitudes $\hat{G}^{ab}_{\mu\nu}$
and $\hat{G}_l^{ij}$ with external gluonic or quarkonic states respectively (it turns out that external ghost states always lead to vanishing traces).
The gluonic projector is given by
%---------------------------------------------------------------------------------
\begin{eqnarray}
\label{eq:Pg}
P_g \hat{G}^{ab}_{\mu\nu} = \frac{\delta^{ab}}{N_C^2-1} \frac{1}{(D-2)(D-3)} (\Delta p)^{-N-1} 
\ep^{\mu\nu\rho\sigma} 
\Delta_\rho p_\sigma \hat{G}^{ab}_{\mu\nu},
\end{eqnarray}
%---------------------------------------------------------------------------------
while we use 
%---------------------------------------------------------------------------------
\begin{eqnarray}
\label{eq:PqNEW}
P_q \hat{G}_l^{ij} = - \delta_{ij} \frac{i (\Delta.p)^{-N-1}}{4 N_C (D-2)(D-3)} \ep_{\mu \nu p \Delta} {\rm tr} \left[p 
\hspace*{-2mm} 
\slash 
\gamma^\mu \gamma^\nu \hat{G}_l^{ij}\right]
\end{eqnarray}
%---------------------------------------------------------------------------------
for external light quarks.
This quarkonic projector differs from the one proposed in Ref.~\cite{Buza:1996xr}
but yields the proper definition of the massive OMEs with massless 
external quark lines in the Larin scheme, unlike the one in Ref.~\cite{Buza:1996xr}.
For more details see \cite{Behring:2019tus,POL19}.
The existence of a single 
projector (\ref{eq:PqNEW}) allows the use of the methods already developed for the unpolarized case. 
This also applies to the calculation of a series of fixed moments using {\tt MATAD} 
\cite{Steinhauser:2000ry}. 

The $O(\ep^0)$ and $O(\ep)$ contributions to the 2-loop
OMEs are needed for the renormalization.
They are given in 
\cite{POL19} for $a_{Qg}^{(2)}$, \cite{POL19,Blumlein:2019zux} for $a_{qq,Q}^{(2),\rm PS}$,
\cite{Hasselhuhn:2013swa} for $a_{gg,Q}^{(2)}$, \cite{PVFNS} for $a_{gq,Q}^{(2)}$.
Although the non-singlet OMEs can be given in the $\MS$ scheme without a recomputation, 
it has to be supplied in the Larin-scheme for the consistent calculation in one scheme.
The contributions are given by
%------------------------------------------------------------------------------------------------------------------------
\begin{eqnarray}
a_{qq,Q}^{(2),\rm NS} &=& \textcolor{blue}{C_F T_F} \Biggl\{ 
\frac{R_1}{54 N^3 (N+1)^3} 
+ \Biggl(\frac{2 (2 + 3 N + 3 N^2)}{3 N ( N+1)} - \frac{8}{3} S_1 \Biggr) \zeta_2 - 
    \frac{224}{27} S_1 + \frac{40}{9} S_2 - \frac{8}{3} S_3 \Biggr\}
\nonumber\\
\\
\overline{a}_{qq,Q}^{(2),\rm NS} &=& \textcolor{blue}{C_F T_F} \Biggl\{ 
\frac{R_2}{648 N^4 (1 + N)^4} + 
      \Biggl(\frac{2 (2 + 3 N + 3 N^2)}{9 N ( N+1)} - \frac{8}{9} S_1 \Biggr) \zeta_3
    + \Biggl(\frac{R_3}{18 N^2 (N+1)^2} 
\nonumber\\ &&
- 
       \frac{20}{9} S_1 
+ \frac{4}{3} S_2\Biggr) \zeta_2 
- \frac{656}{81} S_1 + \frac{112}{27} S_2 - \frac{20}{9} S_3 + \frac{4}{3} S_4
\Biggr\},
\\
R_1 &=& 72 + 240 N + 344 N^2 + 379 N^3 + 713 N^4 + 657 N^5 + 219 N^6,
\\
R_2 &=& -432 - 1872 N - 3504 N^2 - 3280 N^3 + 1407 N^4 + 7500 N^5 + 9962 N^6 + 6204 N^7 
\nonumber\\ &&
+ 1551 N^8,
\\
R_3 &=& -12 - 28 N - N^2 + 6 N^3 + 3 N^4.
\end{eqnarray}
%------------------------------------------------------------------------------------------------------------------------

%--------------------------------------------------------------------------------------------------------
\section{Calculation Methods and Results} 
\label{sec:3}
%--------------------------------------------------------------------------------------------------------

\vspace*{1mm}
\noindent
For the calculation of the pole parts of the OMEs $A_{qq,Q}^{(3),\rm PS}, A_{Qq}^{(3),\rm PS}, A_{qg,Q}^{(2)}, A_{gq,Q}^{(3)}$ and
$A_{gg,Q}^{(3)}$ we were able to use standard techniques like
the method of hypergeometric functions \cite{HYP,SLATER}, the method of hyperlogarithms
\cite{Brown:2008um,Ablinger:2014yaa,Panzer:2014caa}, the solution of ordinary differential equation systems 
\cite{DEQ,Ablinger:2015tua,Ablinger:2018zwz} and the Almkvist--Zeilberger algorithm \cite{AZ,Ablinger:PhDThesis}, see \cite{Blumlein:2018cms} for a survey of these methods, since no elliptic integrals
contribute even in higher orders in the dimensional regulator $\ep$.
The master integrals necessary for the calculation have been already available 
from the calculation of the unpolarized three--loop anomalous dimensions in Ref.~\cite{Ablinger:2017tan}, only a few additional integrals had to be solved
using the method of differential equations.
In all of the above methods corresponding sum 
representations have been derived which were solved using the difference--field techniques 
\cite{Karr:81,Schneider:01,Schneider:05a,Schneider:07d,Schneider:10b,Schneider:10c,Schneider:15a,Schneider:08c} 
of the packages {\tt Sigma} \cite{SIG1,SIG2}, {\tt EvaluateMultiSums}, {\tt SumProduction} \cite{EMSSP}, and using
{\tt HarmonicSums} \cite{Vermaseren:1998uu,Blumlein:1998if,HARMONICSUMS,Ablinger:PhDThesis,Ablinger:2011te,Ablinger:2013cf,
Ablinger:2014bra}.

These methods however do not work for the OME $A_{Qg}^{(3)}$. 
Using standard techniques one encounters elliptic contributions 
\cite{Ablinger:2017bjx,Blumlein:2018aeq}, which cannot be handled 
in the same automated way.
We therefore apply the method of arbitrarily 
large moments \cite{Blumlein:2017dxp} in this case.
Using this method one can recursively generate higher and higher
moments of the master integrals and thereby the complete OME.
These moments are used to derive a difference equation by the method of guessing \cite{GUESS} implemented in {\tt Sage}
\cite{SAGE,GSAGE}.

Due to the structure of the IBP relations some higher expansion in 
$\ep$ is 
necessary also to extract the term $\propto 1/\ep$. Here one would encounter  elliptic terms by using the above 
techniques. We therefore apply the method of arbitrarily 
large moments \cite{Blumlein:2017dxp} in this case.\footnote{This method has been successfully applied also in a series 
of other calculations, cf.~\cite{Ablinger:2017tan,GUESS1,Blumlein:2009tj}.} Here one works in moment--space and the IBP 
relations
are expressed in terms of recurrences for the master integrals. Using these relations one generates 
systematically higher and higher moments both for the master integrals and the operator matrix elements.
We generated 2000 Mellin moments, which allowed to find most of the recurrences for all seventeen color--$\zeta$ 
projections. To determine the recurrences of the projections $C_F C_A T_F$ and $C_A^2 T_F$ we used 4000 moments, out 
of which 2640 turned out to be sufficient. 
Here we refer to representations in terms of even and odd moments, with 
the even moments being unphysical. The analytic continuation is finally performed from the odd moments only.
The characteristics of the recurrences for the different color--$\zeta$ factors contributing to the $1/\ep$ term of 
the unrenormalized massive OME $A_{Qg}^{(3)}$ are summarized in Table~1. For all the pole terms these 
recurrences are first--order factorizable and can be solved by applying the package {\tt Sigma}.
Here some color--$\zeta$ structures contribute for technical reasons, which cancel in the final expression.

All anomalous dimensions can be expressed by nested harmonic sums \cite{Vermaseren:1998uu,Blumlein:1998if}
%------------------------------------------------------------------------------------------------------------------------
\begin{eqnarray}
S_{b,\vec{a}}(N) &=& \sum_{k=1}^N \frac{({\rm sign}(b))^k}{k^{|b|}} S_{\vec{a}}(k),~~~S_\emptyset = 1~~~,b, a_i \in
\mathbb{Z} \backslash \{0\}. 
\end{eqnarray}
%------------------------------------------------------------------------------------------------------------------------
To provide comparisons on a diagram-by-diagram basis we have calculated the first few Mellin moments for $N = 1, 3, 
5, 7, 9$
using {\tt MATAD} \cite{Steinhauser:2000ry}.

We would like to compare to the results obtained in Ref.~\cite{Moch:2014sna} which are given in the
M--scheme. This scheme was defined in implicit form in Ref.~\cite{Matiounine:1998re}. Up to two--loop order 
it is the same as the one in which the results of Refs.~\cite{Mertig:1995ny,SP_PS1} were obtained.
At leading order, the anomalous dimensions are scheme--invariant.
The finite renormalizations between the Larin and the M--scheme to three--loop order can be obtained following 
\cite{Matiounine:1998re}, see also \cite{Moch:2014sna}. 
For the explicit transformations and the results on the complete polarized 2--loop and $T_F$-contributions to the 
polarized 3--loop anomalous dimensions, see \cite{Behring:2019tus}. At 3--loop order the complete anomalous dimensions 
$\gamma_{qq}^{(2),\rm PS}$ and $\gamma_{qg}^{(2)}$ are obtained.
Ref.~\cite{Behring:2019tus} also contains the results for the anomalous dimensions and splitting
in computer-readable form. We fully agree with the results given in \cite{Moch:2014sna}.

\begin{table}[h!]
\centering
\def\arraystretch{1.5}% 
\begin{tabular}{|l|r|r|}
\hline
\multicolumn{1}{|c}{color/$\zeta$} & \multicolumn{1}{|c}{order} & \multicolumn{1}{|c|}{degree}\\  
\hline
$C_F T_F^2$                &  7  &  68    \\
$C_F T_F^2 \zeta_2$        &  3  &  17    \\
$C_F T_F^2 N_F$            &  7  &  68    \\
$C_F T_F^2 N_F \zeta_2$    &  3  &  17    \\
$C_F^2 T_F$                & 22  & 283    \\
$C_F^2 T_F \zeta_2$        &  6  &  32    \\
$C_F^2 T_F \zeta_3$        &  2  &  10    \\
$C_A T_F^2$                & 10  &  85    \\
$C_A T_F^2 \zeta_2$        &  3  &  12    \\
$C_A T_F^2 N_F$            & 14  & 131    \\
$C_A T_F^2 N_F \zeta_2$    &  4  &  16    \\
$C_F C_A T_F$              & 30  & 484    \\
$C_F C_A T_F \zeta_2$      &  8  &  46    \\
$C_F C_A T_F \zeta_3$      &  3  &  19    \\
$C_A^2 T_F$                & 30  & 472    \\
$C_A^2 T_F \zeta_2$        & 10  &  57    \\
$C_A^2 T_F \zeta_3$        &  4  &  19    \\
\hline
\end{tabular}
\def\arraystretch{1}% 
\label{TAB1}
\caption{Characteristics of the recurrences contributing to the anomalous dimension $\gamma_{qg}^{(2)}$.}
\end{table}

%--------------------------------------------------------------------------------------------------------
\section{Conclusions} 
\label{sec:4}
%--------------------------------------------------------------------------------------------------------

\vspace*{1mm}
\noindent
Since the QCD anomalous dimensions are universal quantities one can compute
them within various setups.
In the present case they have been obtained from the pole structure 
of massive polarized OMEs at three loop order.
The calculation of these OMEs is part of an ongoing project with the final goal to compute the 
massive polarized Wilson coefficients for deep--inelastic scattering in the region $Q^2 \gg m^2$.
Through this calculation we got the contributions $\propto T_F$ to the polarized 3--loop anomalous dimension 
$\gamma_{ij}^{(2)}(N)$ and the associated splitting functions in a massive calculation.
This calculation is fully independent of the earlier computation in Ref.~\cite{Moch:2014sna}. 
We completely agree with the previous results. 
To use conventional IBP reduction techniques we resum the local operator
into a formal Taylor series using the auxiliary parameter $x$.
As in the unpolarized case \cite{Ablinger:2017tan} before, we had to use the method of arbitrarily 
high moments \cite{Blumlein:2017dxp} to deal with potential elliptic contributions in the necessary deeper 
expansions in the dimensional parameter $\ep$ in the case of the OME $A_{Qg}^{(3)}$. 
Using the method of high Mellin moments \cite{Blumlein:2017dxp}, the moments  
are calculated recursively using the system of difference equation associated with the differential
equations given by the IBP relations. Individual master integrals are only calculated in terms of moments. 
In all other contributions, standard techniques, cf.~\cite{Blumlein:2018cms}, are used in the calculation of the 
master integrals.
%------------------------------------------------------------------------------------------------------------------
\section*{Acknowledgments}
%------------------------------------------------------------------------------------------------------------------

\noindent 
We thank J.~Gracey, F.~Herren, S.~Moch, D.~St\"ockinger, A.~Vogt and S.~Weinzierl 
for discussions. This work has been funded in part by EU TMR network SAGEX agreement No. 764850
(Marie Sk\l{}odowska-Curie) and COST action CA16201: Unraveling new physics at
the LHC through the precision frontier. 
%--------------------------------------------------------------------------------------------------------
%-----------------------------------------------------------------------------------------------------

%-----------------------------------------------------------------------------------------------------
%-----------------------------------------------------------------------------------------------------

\end{document}